\documentclass[11pt,a4paper]{article}
\usepackage{graphicx}
\usepackage{amssymb}
\usepackage{epsfig}
\usepackage{fullpage}
\newcommand{\bm}[1]{\mbox{\boldmath$#1$}}
\title{Emission spectrum of the atomic chain excited by channeled particle}
\author{V. Epp, M. A. Sosedova\\
{\it Tomsk State Pedagogical University, Tomsk, 634061, Russia}
}
\date{}
\begin{document}
\def\bea{\begin{eqnarray}}
\def\eea{\end{eqnarray}}
\def\nn{\nonumber}
\maketitle


\begin{abstract}
Basic properties of radiation of the atomic chains excited by a channeled particle are considered. Using a very simple two-dimensional model of a crystal lattice we have shown that the main part of this radiation is generated on the frequency of oscillations of a channeled particle between the crystal planes, shifted by the Doppler effect. Spectral and angular distribution and spectral distribution of the radiation of the atomic chain excited by channeled particle were calculated. Emission spectrum of the atomic chain excited by channeled particle was plotted.
\end{abstract}





\section{Introduction}

The motion of charged particles in the crystal along the channels formed by parallel atomic rows or planes was predicted by American scientists M.T. Robinson and O.S. Oen in 1961 \cite{Robin} and was discovered by several groups of scientists in 1963 -- 1965. Channeling has served as the base for development of new experimental methods of research on the crystal properties and structure and some nuclear phenomena. New sources of X-ray and gamma rays which consist of electrons or positrons accelerators and precisely oriented crystals of germanium, silicon and other elements are based on the principles of channeling. 

Channeling of ions is used recently for direct detections of weakly interacting massive particles \cite{Drob, Berna}. Considerably attention has been attracted currently to study the carbon nanotubes properties by use of channeling.

At interaction between the channeled particle and a crystal the electromagnetic radiation is generated. In the framework of classical electrodynamics one can consider this radiation as radiation from different sources, such as radiation of the channeled particle, which is investigated in details in many papers and books \cite{Lindhard, Baz}, radiation of the electronic gas (wake trace) \cite{Kil} and the less studied radiation of excited atoms in the crystal lattice. 

When channeling, the charged particle transmits part of its transverse momentum to the surrounding atoms. This excites vibration of atomic chains. Oscillations of a nucleus with interior electrons cause time-dependent polarization of atom and hence electromagnetic radiation. This phenomenon is similar to the vibrational excitation of molecules and causes respective radiation. The phases of oscillations of atoms correlate between themselves as oscillations are excited by the same channeled particle. Therefore radiation should be considered as coherent.

Assumption of such radiation was expressed recently in \cite{Sosedova}. It was shown that the basic part of radiation is generated into a cone around the direction of velocity of a relativistic channeling particle. Correlation between the fields of atomic radiation and radiation of the channeled particle was discussed in \cite{NIM_B}.

In this paper, we study the field of oscillating atoms in detail and calculate the emission spectrum of the atomic chains, excited by a channeled particle. In section \ref{sec2} we  discuss some of the features of the radiation field of the atomic chains, which were not taken into account in \cite{Sosedova}. The next section presents the calculation of the spectral and angular distribution of the radiation of the atomic chain. The last section \ref{sec4} is devoted to the calculation and construction of the emission spectrum of atomic chain.

\section{Field of atomic chains excited by a channeled particle} \label{sec2}

\begin{figure}[tbh!]
\centerline{\includegraphics[width=3in]{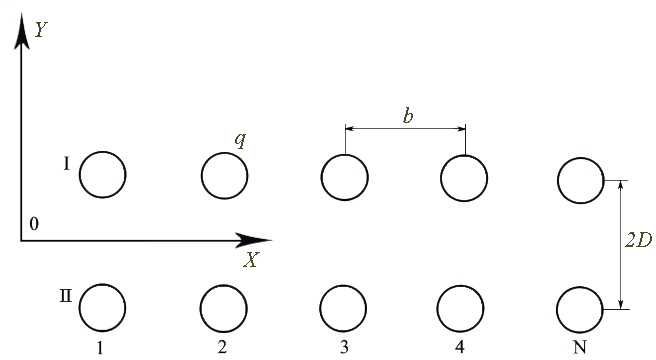}}\caption{Schematic representation of the crystal lattice.} \label{fig1}
\end{figure}

\begin{figure}[tbh!]
\centerline{\includegraphics[width=2.5in]{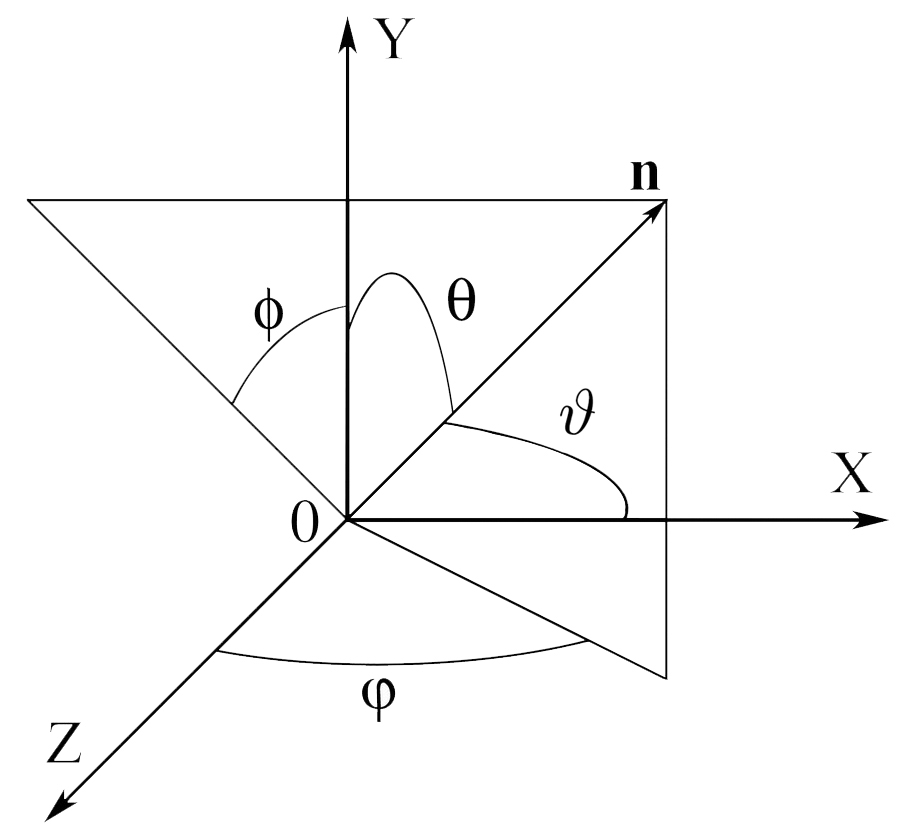}}\caption{The spherical coordinate system.}
\label{fig2}
\end{figure}

Here we consider the electric field of the radiation of the oscillating atoms in a crystal. Following \cite{Sosedova}, we consider the simplest model of the crystal lattice, consisting of two atomic rows with a channeled particle moving between the rows. Of course, such a model is a rather crude approximation of the real crystal, but it allows to identify the characteristics of the phenomenon. We are encouraged by the fact that in due time the normal modes of vibrations in the crystal were investigated by use of one-dimensional crystal lattice. Two-dimensional model of the crystal lattice is shown in figure \ref{fig1}. 
The electric field $E(t)$ of the radiation of oscillating atoms of lattice is described by the equation (33) in \cite{Sosedova}. Let us rewrite this equation in the complex form
\bea\label{eott2}
E(t)=A_0\nu^2\left\{\frac{e^{i\nu t}- e^{- i \Omega 't}}{\nu+\Omega'}+\frac{e^{i\nu t}- e^{i\Omega 't}}{\nu-\Omega'}\right\},
\eea
for $t\leq Lk$ and
\bea\label{eott3}
E(t)=A_0\nu^2 e^{i\nu t}\left\{\frac{1-e^{-i kL(\nu+\Omega')}}{\nu+\Omega'}+\frac{1-e^{-i kL(\nu-\Omega')}}{\nu-\Omega'}\right\},
\eea
for $t> Lk$. Here 
\bea \label{amplit}
A_0 =\frac {eq^2K\sin\theta} {kmVD^2\omega bRc^2},\quad \Omega'=\frac{\Omega}{(1-\beta\sin\theta\sin\varphi)},\\
 k=\frac {1}{V} (1-\beta\sin\theta\sin\varphi),\quad \nu=\omega+i\alpha.
\eea
$V$ is the velocity of  channeled particle, $\beta=V/c$, $c$ is the speed of light, $e$ is the charge of the channeled particle, $m$ is the mass of the atom, $2D$ is the width of the channel, $b$ is the distance between atoms in the rows, $q$ is an effective charge of a nucleus of atom, $L$ is the length of the crystal along the particle trajectory. $R$ is the distance to the observer. Angles $\theta$ and $\varphi$  of the spherical coordinate system are shown in figure \ref{fig2}. $K$ and $\Omega$ are the amplitude and frequency of channeling particle oscillations in plane $XOY$. The average velocity of a particle is directed along the axis $X$. The particle enters the crystal at $x=0$ and leave it at $x=L$. The excited atoms of the crystal lattice are oscillating along axis $Y$ with frequency $\omega$. The amplitude of vibrations decreases with time as $\exp (-\alpha t)$, where $\alpha$ is the attenuation coefficient. In the spherical coordinates vector of the radiation field $\bm E(t)$  has only one component: $E_{\theta}=E(t)$.
We see from  (\ref{eott2}) and (\ref{eott3})  that radiation is generated on two frequencies, namely on frequency $\omega$ of atoms oscillations  and on the frequency $\Omega'$ which is frequency of oscillations of the channeling particle shifted by the Doppler effect. 
The part of the field oscillating with frequency $\omega$ vanishes exponentially in time as $e^{-\alpha t}$, while the field of frequency $\Omega'$ is constant in amplitude. In case of long enough crystal, when the length of a crystal is much greater then the ``wake tail'' of atoms exited by the channeling particle ($L \gg V\alpha^{-1}$), the part of radiation with frequency $\Omega'$ makes much greater contribution to the total energy of radiation.

Further we consider only this part of radiation. Taking the real part of equation (\ref{eott2}) we obtain

\bea\label{eott5}
E(t)&=& -\frac{A_0\omega}{2\left[(\alpha^2+\omega^2-\Omega'^2)^2+4\alpha^2\Omega'^2\right]}\nn\\
&\times&\left\{2\alpha\Omega'^3\sin\Omega't
+\cos\Omega't\left[\left(\alpha^2 +\omega^2\right)^2\right.\right.\nn\\
&+&\left.\left.\Omega'^2(3\alpha^2-\omega^2)\right]\right\}.
\eea

This field has some features typical for the field of relativistic charged particle though the radiation is caused by non-relativistic vibrating atoms. First, the multiple $1-\beta\sin\theta\sin\varphi$ in denominator of $A_0$ shows that the main part of radiation is concentrated around the direction of the particle velocity $\bm \beta$.  Second, the frequency of radiation $\Omega'$ is just the same as the frequency of the relativistic channeling particle. Coherence of radiation is expressed by the factor $b$ in denominator of $A_0$. It shows that intensity of radiation is proportional to the square of  linear charge density $q/b$. Denominator of (\ref{eott5}) is of typical resonant character. Indeed, if attenuation of atomic vibrations vanishes ($\alpha\to0$) and $\Omega'=\omega$, the electric field defined by the above equation tends to infinity.
\section{Spectral and angular distribution of the radiation of atomic chain} \label{sec3}
Let us calculate spectral and angular distribution of the radiation field (\ref{eott2}) and (\ref{eott3}) of the atoms. We find the Fourier integral 
\bea\label{furie1}
E(\omega_1)=\int\limits_{-\infty}^\infty e^{i\omega_1 t}E_{\rm r}(t)\rm dt,
\eea
where $\omega_1$ is the frequency of radiation, $E_{\rm r}(t)=\displaystyle\frac{1}{2}[E(t)+E^*(t)]$ is the real part of the field (\ref{eott2}) and (\ref{eott3}). In order to simplify the result of integration, we assume that the channeled particle makes an integer number of oscillations $N$ in crystal. Then $L=2\pi NV/\Omega$, and some  exponents can be simplified: $\exp(\pm i Lk\Omega')=1.$
The result of integration looks like follows:
\bea\label{furie0}
E(\omega_1)=\frac{2A_0 i \omega\omega_1(1-e^{i \omega_1 Lk})(\alpha^2+\omega^2-2 i\alpha\omega_1)}{(\omega_1^2-\Omega'^2)(\omega_1^2-\omega^2-\alpha^2+2 i \alpha\omega_1)}.
\eea

Spectral and angular distribution of energy $\rm d{\cal E}$ emitted into solid angle $\rm do$ and frequency interval $\rm d\omega_1$  is defined by the formula \cite{Landau}:
\bea\label{spugl}
\frac{\rm d{\cal E}}{\rm do \rm d\omega_1}=\frac{cR^2}{4\pi^2}\left|E(\omega_1)\right|^2.
\eea
Taking the square of module $E(\omega_1)$ and  substituting it in the last equation we obtain:
\bea \label{sp-ug}
\frac{\rm d{\cal E}}{\rm do \rm d\omega_1}={\cal E}_0I_1(\widetilde\omega,\theta,\varphi)I_2(\widetilde\omega),
\eea
where ${\cal E}_0$ is a constant, and ``tilde'' denotes the reduced variables:
\[
{\cal E}_0=\frac {16e^2q^4K^2\gamma^2}{\pi^2m^2c^3D^4b^2\omega^2},\,\, \widetilde{\omega}=\displaystyle\frac{\omega_1}{\omega},\,\, \widetilde{\alpha}=\frac{\alpha}{\omega},\,\, \widetilde{\Omega'}=\frac{\Omega'}{\omega}.
\]
Function $I_1(\widetilde\omega,\theta,\varphi)$ depends on the angles and parameters of the channelled particle
\bea\label{I_1nonrel}
I_1(\widetilde\omega,\theta,\varphi)=\frac{\widetilde\omega^2\sin^2\theta\sin^2(\pi N\widetilde\omega/\widetilde\Omega')}
{4\gamma^4(1-\beta\sin\theta\sin\varphi)^2(\widetilde\omega^2-\widetilde\Omega'^2)^2}.
\eea
In case of large $N$ this function  demonstrates the resonant amplification of coherent atomic vibrations at the reduced frequency $\widetilde\omega=\widetilde\Omega'$.

Function $I_2(\widetilde\omega)$ depends only on the reduced frequency. It has sharp maximum at $\widetilde\omega=1$ if  $\alpha$ is small enough
\bea\label{funI_2}
I_2(\widetilde\omega)=\frac{(1+\widetilde\alpha^2)^2+4\widetilde\alpha^2\widetilde\omega^2}{[(\widetilde\omega^2-\widetilde\alpha^2-1)^2+4\widetilde\alpha^2\widetilde\omega^2]}.
\eea

The multiplier $(1-\beta\sin\theta\sin\varphi)$ in denominator of (\ref{I_1nonrel}) shows that in case of ultrarelativistic channelled particle,  the function $I_1$ is relative great at small angles around the direction of the velocity. In ultrarelativistic case $\gamma\gg 1$ it is more convenient to describe the angular distribution in terms of angles $\vartheta$ and $\phi$ (see figure \ref{fig2}) which depend on $\theta$ and $\varphi$ as follows:
\bea\label{newAngl}
\sin\theta\sin\varphi=\cos\vartheta,\,\,  \sin^2\theta=\cos^2\phi+\cos^2\vartheta\sin^2\phi.
\eea
We expand $I_1$ with respect to small $\vartheta$ and $\gamma$ and denote $\psi=\gamma\vartheta$. Then we have instead of (\ref{I_1nonrel})
\bea\label{I_1rel}
I_1(\psi,\widetilde\omega)=\frac{\widetilde\omega^2\sin^2(\pi N\widetilde\omega/\widetilde\Omega')}
{(1+\psi^2)^2(\widetilde\omega^2-\widetilde\Omega'^2)^2},
\eea
with 
\[
\widetilde\Omega'=\frac{2\gamma^2\widetilde\Omega}{1+\psi^2},\quad \widetilde\Omega=\frac{\Omega}{\omega}.
\]

\begin{figure}[tbh!]
\centerline{\includegraphics[width=4in]{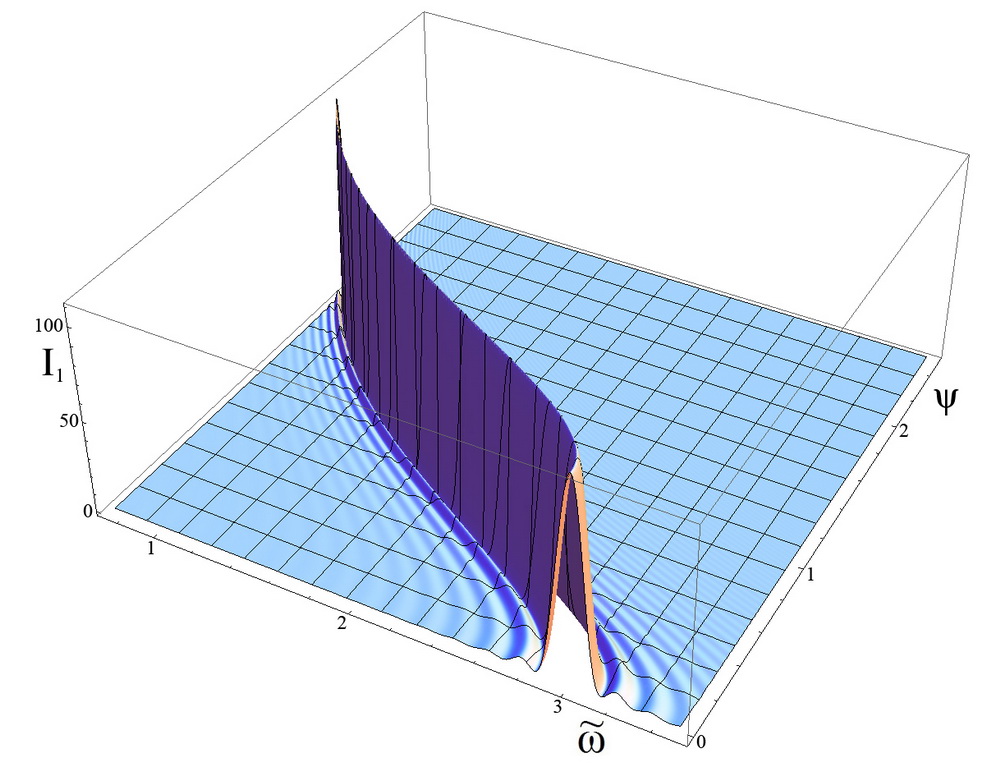}}\caption{Function $I_1(\widetilde\omega,\psi)$.} \label{I_1}
\end{figure}

Figure \ref{I_1} shows behaviour of  function $I_1(\psi,\widetilde\omega)$  at  the following parameters: $N=20$, $\widetilde{\Omega}=1.5\times10^{-4}, \gamma=100$.

The plot constitutes a ridge laying above the line $\widetilde\omega=2\gamma^2\widetilde\Omega/(1+\psi^2)$ in the plane  ($\widetilde\omega, \psi$). The width of the ridge is inversely related to the number of oscillations $N$ and the maximum of $I_1$ is in direct proportion to $N$.

\begin{figure}[tbh!]
\centerline{\includegraphics[width=4in]{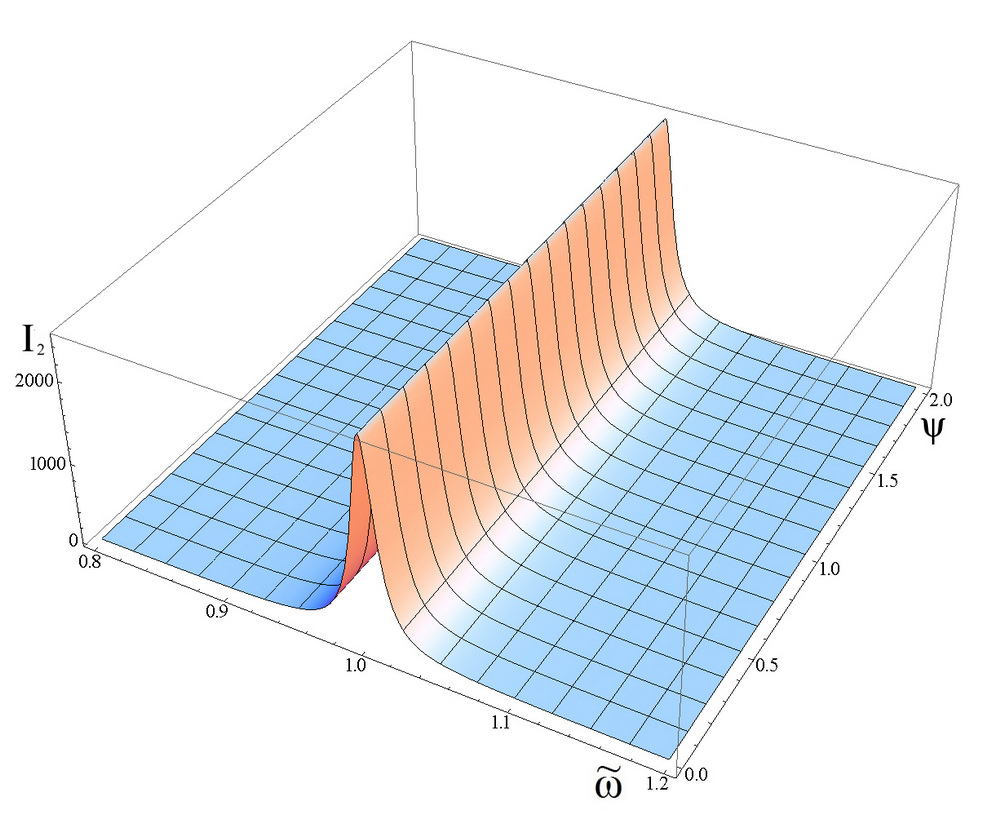}}\caption{ Function $I_2(\widetilde\omega)$.} \label{I_2}
\end{figure}

Figure \ref{I_2} represents the function $I_2(\widetilde\omega)$ for  $\widetilde{\alpha}=10^{-2}$
in the some coordinates. It follows from (\ref{funI_2}) that in case of small $\alpha$ the maximum of the function is inversely  proportional to $\alpha^2$ and the width of the peak is proportional to $\alpha$.

\begin{figure}[tbh!]
\centerline{\includegraphics[width=4in]{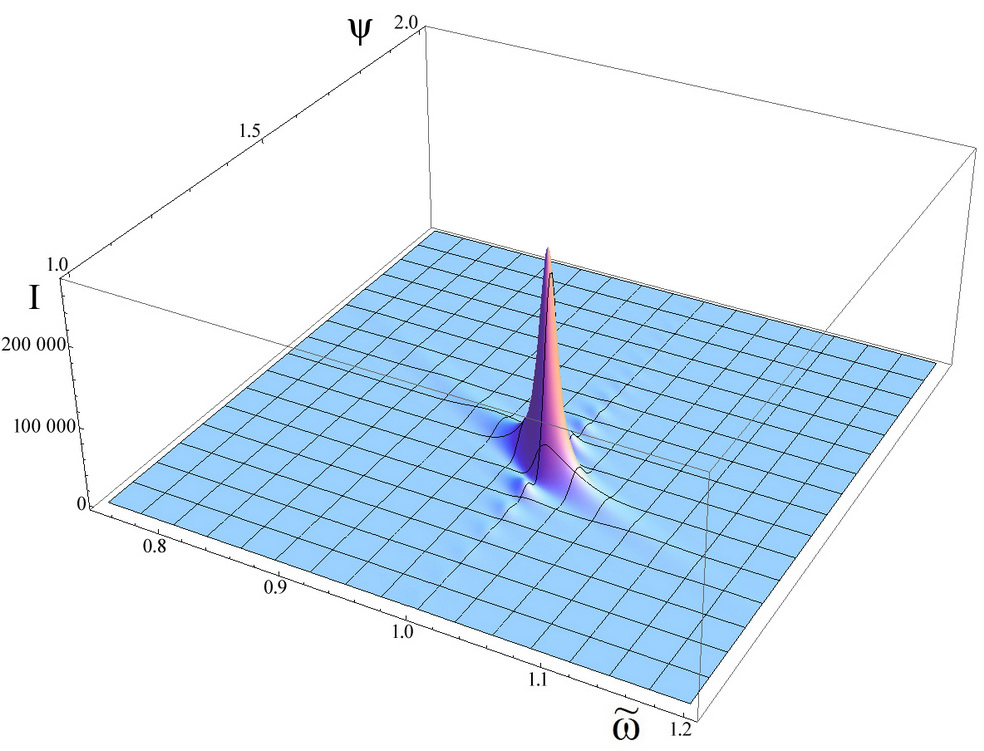}}\caption{ Function$I(\psi,\widetilde\omega);\,\, 2\gamma^2\widetilde\Omega=3$.} \label{intersec}
\end{figure}

Figure \ref{intersec} depicts the  product $I(\psi,\widetilde\omega)=I_1I_2$ which (according to (\ref{sp-ug}))  is up to a scaling factor the energy radiated per unit frequency per unit solid angle. The narrow strong pick lies at intersection of lines $\widetilde\omega=1$ and $\widetilde\omega=2\gamma^2\widetilde\Omega/(1+\psi^2)$ in the plane  ($\widetilde\omega, \psi$). Such intersection is possible only if $2\gamma^2\widetilde\Omega>1$.

\begin{figure}[tbh!]
\centerline{\includegraphics[width=4in]{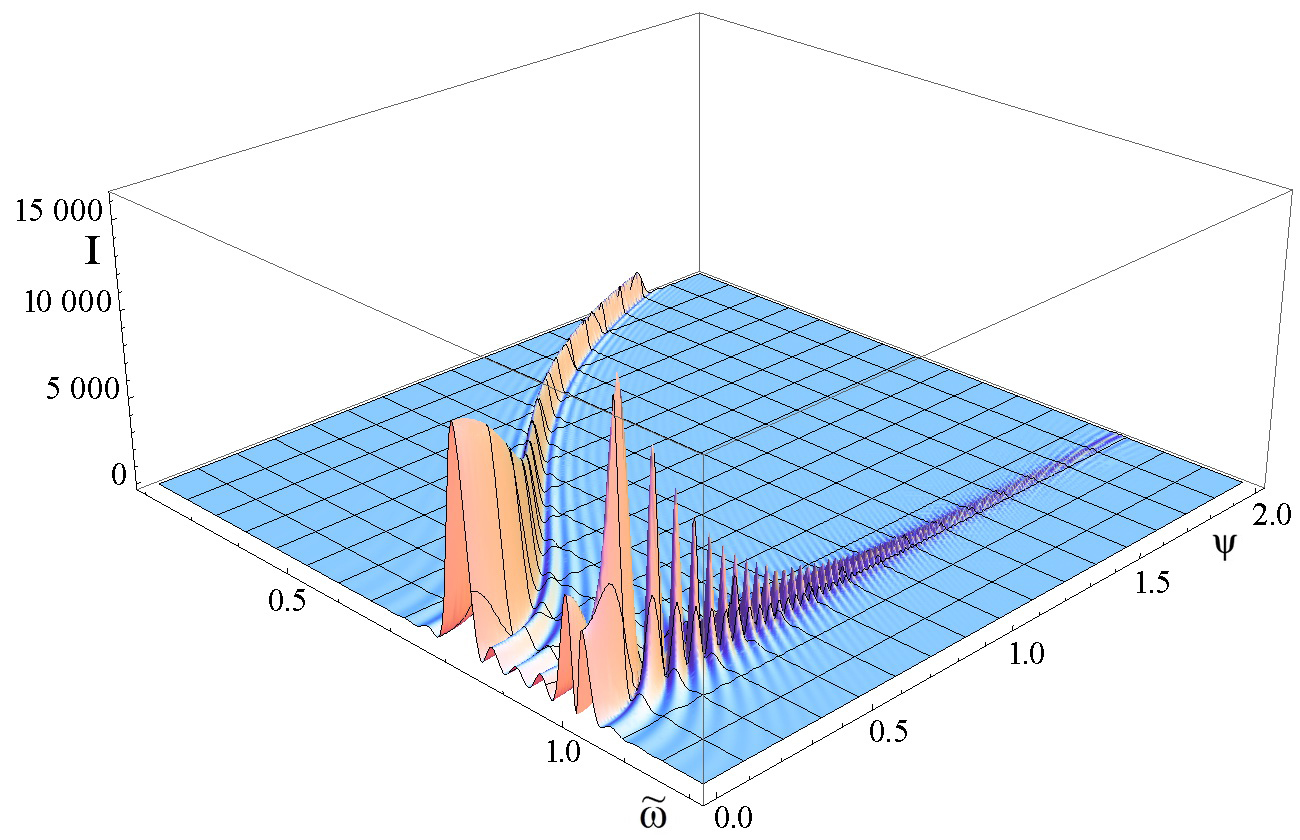}}\caption{ Function $I(\psi,\widetilde\omega);\,\, 2\gamma^2\widetilde\Omega=0.8$.} \label{intersec1}
\end{figure}

 In case $2\gamma^2\widetilde\Omega<1$ the function $I(\psi,\widetilde\omega)$ has complicated structure  as shown in figure~\ref{intersec1}. The maximum value in this case is about one order of magnitude less than in figure~\ref{intersec}.

Let us calculate the angular distribution. Integration of  (\ref{sp-ug}) over frequency $\omega_1$ can be simplified in case of large $N$. It is easy to check that the function $f(x)$
\[
f(x)=\lim_{N\to\infty}\frac{\sin^2\pi Nx}{N(1-x^2)^2}
\]
possess next properties
\[
f(x)=\cases{	 \infty,  & $x=\pm 1$,
\cr
0,  & $x\neq \pm1$}, \; \int\limits_{-\infty}^\infty f(x){\rm d }x=\frac 12\pi^2.
\]
Hence, it can be replaced by sum of the delta functions
\begin{equation}\label{d_func}
f(x)=\frac{\pi^2}{2}[\delta(x-1)+\delta(x+1)].
\end{equation}
Integrating  (\ref{sp-ug}) under assumption that $N$ is great, we obtain the angular distribution of energy, emitted by the atomic chain
\bea\label{ugl}
\frac{\rm d{\cal E}}{\rm do}&=&\frac{e^2q^4K^2N\sin^2\theta}{m^2c^3D^4b^2\Omega'(1-\beta \sin\theta\sin\varphi)^2}\nn\\
&\times&\frac{(\omega^2+\alpha^2)^2+4\alpha^2\Omega'^2}{[(\Omega'^2-\alpha^2-\omega^2)^2+4\alpha^2\Omega'^2]}.
\eea

Expression (\ref{ugl}) coincides with (35) in \cite{Sosedova} for the radiation intensity, because the intensity and energy of radiation are bound by equation $\rm d I=\Omega'\rm d{\cal E}/2\pi N$ \cite{Bordovitsyn}.
\section{Emission spectrum of the atomic chain} \label{sec4}
In order to calculate the emission spectrum we integrate (\ref{sp-ug}) over the solid angle $do$. Integration is better done in the spherical coordinates  system $(\vartheta,\phi$) (see (\ref{newAngl}) and figure~\ref{fig2}).

We denote the ratio $\omega_1/\Omega'\equiv\xi$. Integral over the solid angle $\rm do=\sin\vartheta \rm d\vartheta \rm d\phi$ can be expressed as integral over $\phi$ and $\xi$. The limits of integration over $\xi$ are equal to:
\bea\label{lim}
\xi_1\equiv\displaystyle\frac{\omega_1}{\Omega}(1-\beta), \quad \xi_2\equiv\displaystyle\frac{\omega_1}{\Omega}(1+\beta).
\eea

Again, we use the approximation of great $N$ and replace $\sin^2[\pi N\xi]/(\xi^2-1)^2$  by the sum of  $\delta$ functions according to  (\ref{d_func}).
The expression $\rm d{\cal E}(\omega_1)/\rm d\omega_1\neq0$, if a $\delta$ function is inside the limits of integration (\ref{lim}). It means that the radiation lies in the frequency range
\[
\frac{\Omega}{1+\beta}<\omega_1<\frac{\Omega}{1-\beta}.
\]
Integrating (\ref{sp-ug}), we obtain the final version for the emission spectrum of the atomic chain:

\bea\label{spectrum2}
\frac{\rm d{\cal E}}{\rm d\widetilde{\omega}}=\frac{2e^2q^4K^2\pi N}{m^2c^3D^4b^2\Omega\omega}\frac{\left[\widetilde{\omega}^2(1+\beta^2)+\widetilde{\Omega}^2\right]}{\widetilde{\omega}^3\beta^3}I_2(\widetilde{\omega}).
\eea
The last formula shows that the shape of the emission spectrum does not depend on $\omega_1$ but only on the ratio of  frequencies $\omega_1/\omega$.

\begin{figure}[tbh!]
\centerline{\includegraphics[width=3in]{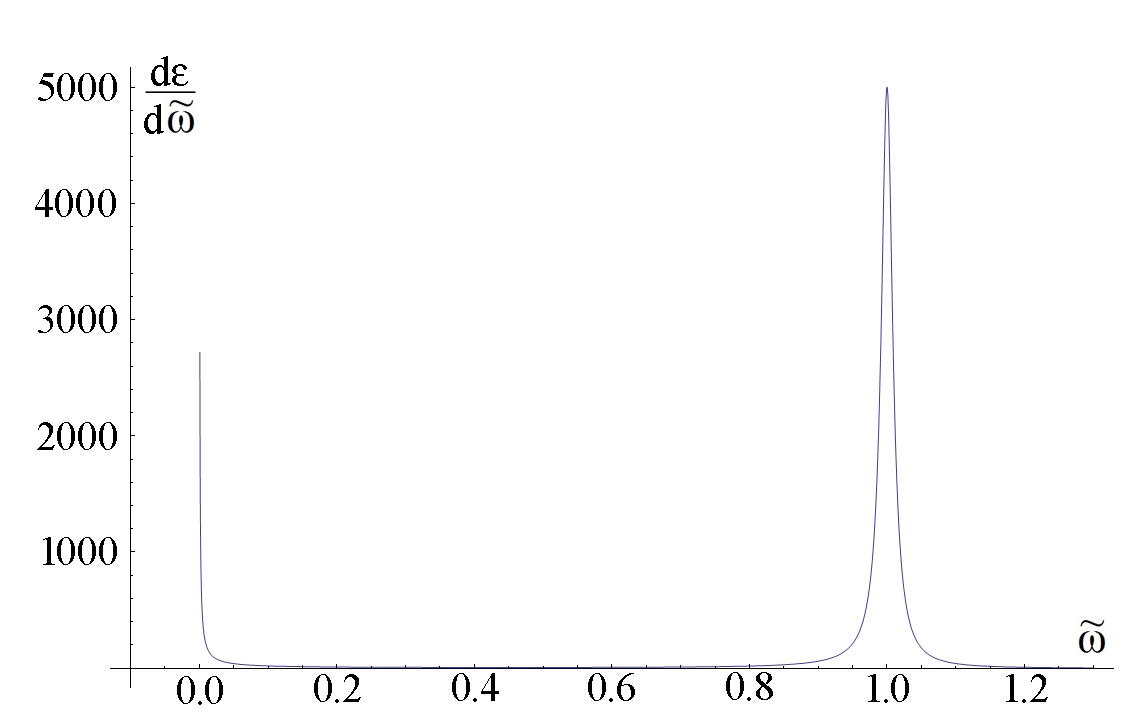}}\caption{Emission spectrum of atomic chain excited by the channeled particle.}
\label{fig3}
\end{figure}

 As an example, we plot the spectrum for the following parameters: $\widetilde{\alpha}=10^{-2}, ~ \widetilde{\Omega}=1.5\times10^{-4}, \gamma=100$. The emission spectrum lies in the range
 \bea
\widetilde{\omega}_{\rm min}=\displaystyle\frac{\widetilde{\Omega}}{1+\beta}\simeq \frac{1}{2}\widetilde{\Omega}=0.75\times10^{-4},\nn\\
\widetilde{\omega}_{\rm max}=\displaystyle\frac{\widetilde{\Omega}}{1-\beta}\simeq 2\widetilde{\Omega}\gamma^2=3.\nn
\eea
Emission spectrum of atomic chain excited by the channeled particle for these parameters is shown in figure \ref{fig3}.
Coherent radiation of atoms, which is discussed in this article expresses a sharp peak at $\widetilde{\omega}=1$. The width of this peak depends on the ratio of  $\alpha/\omega$: the smaller $\widetilde{\alpha}$ the sharper peak in the spectrum. The rise at low frequencies due to  the factor $\widetilde{\omega}^3$ in denominator of the formula (\ref{spectrum2}). Note that $\widetilde{\omega}$ can not be zero, because it is limited to a minimum value $\widetilde{\omega}_{\rm min}=\widetilde{\Omega}/2$.

\section{Conclusion}

At first glance, radiation of atoms exited by a channeled particle and vibrating in the points of crystal lattice is negligible small because of small amplitude and small frequency of vibrations, that is a consequence of relatively great mass of atoms. But if we take into account the great number of atoms which are involved in coherent oscillations excited by the channeled particle, we get sufficient amplification of the intensity of radiation. 
The expressions for the spectral and angular distribution (\ref{sp-ug}) which is obtained in this work shows the specific properties of  considered radiation: concentration of the emitted energy within an narrow cone around the forward direction  if the channeled particle is relativistic; there is a resonance peaking angular distribution  at frequency $\omega_1=\Omega'$ -- the frequency of the radiation of channeled particle; there is also a resonance peak on frequency $\omega_1=\omega$ -- the frequency of vibrations of atoms of the crystal lattice. 

Emission spectrum which is integrated over the angles  has the following features: the boundaries of the spectrum are determined by the frequency of the channeled particles oscillation $\Omega$ -- spectrum lies in the interval $\Omega/(1+\beta)<\omega_1<\Omega/(1-\beta)$; spectrum has the form of a sharp peak at a frequency $\omega_1=\omega$ for small enough values of the attenuation factor $\alpha$. The occurrence of such resonance is possible if  $\Omega/(1+\beta)<\omega<\Omega/(1-\beta)$. 
Or in case of an ultrarelativistic particle $\omega<2\gamma^2\Omega$. The lower boundary of the spectrum is much smaller in this case and can be neglected.

A rather simple model of the crystal lattice has been used just to demonstrate the essence of this phenomenon. Many details have not yet been taken into account.

For example, this paper does not consider the thermal vibrations of the atoms. Thermal vibrations can substantially influence the dynamics of the channeled particle, especially at axial channeling when the negatively charged particle moves close to the atomic chain.
Radiation caused by the thermal vibrations is incoherent and has isotropic angular distribution while the considered radiation of vibrating atoms is coherent and strongly collimated around the radiation cone of the channeled particle. Hence, it can be resolved from the thermal radiation. 

Another factor that has an impact on the coherence of radiation is the number $N$ of oscillations of the channeled particle. In case of thin crystal it depends on the crystal thickness. Otherwise it depends on the length of the particle channelling and is restricted by   de-channelling process owing to multiple scattering  by electrons and thermal vibrations of crystalline lattice atoms. Anyway, if we are speaking about the  channelling of a particle,   we mean that $N\gg 1$.

\section*{Acknowledgement}
This research has been supported by the grant for LRSS, project No 224.2012.2 and by the Ministry of Education and Science of Russian Federation, project 14.B37.21.0774.

\end{document}